\documentclass[aps,psfig,12pt]{revtex4}
\usepackage{epsfig}

\def\beq{\begin{equation}}
\def\eeq{\end{equation}}

\def\bea{\arraycolsep .1em \begin{eqnarray}}
\def\eea{\end{eqnarray}}
\def\Tr{{\rm Tr}}
\def\step{\\[-1.5ex]}

\def\eq#1{(\ref{#1})}

\def\s0#1#2{\mbox{\small{$ \frac{#1}{#2} $}}}
\def\0#1#2{\frac{#1}{#2}}

\def\grgl{\:\hbox to -0.2pt{\lower2.5pt\hbox{$\sim$}\hss}{\raise3pt\hbox{$>$}}\:}
\def\klgl{\:\hbox to -0.2pt{\lower2.5pt\hbox{$\sim$}\hss}{\raise3pt\hbox{$<$}}\:}

\begin{document}


\thispagestyle{empty}
\begin{center}

{\normalsize\begin{flushright}
SHEP 04-31\\
CERN-PH/TH-2005-042\\[15ex] \end{flushright}
}

\mbox{\large \bf Universality and the Renormalisation Group}
 \\[6ex]

{Daniel F. Litim}\\[2ex]
{\it School of Physics and Astronomy\\ 
University of Southampton, Southampton SO17 1BJ, U.K.}
\\[3ex]
{\it Theory Division, CERN, CH-1211 Geneva 23.}
\\[10ex]

{\small \bf Abstract}\\[2ex]
\begin{minipage}{14cm}{\small 
Several functional renormalisation group (RG) equations including Polchinski
    flows and Exact RG flows are compared from a conceptual point of view and
    in given truncations.  Similarities and differences are highlighted with
    special emphasis on stability properties.  The main observations are
    worked out at the example of $O(N)$ symmetric scalar field theories where
    the flows, universal critical exponents and scaling potentials are
    compared within a derivative expansion. To leading order, it is
    established that Polchinski flows and ERG flows -- despite their
    inequivalent derivative expansions -- have {\em identical} universal
    content, if the ERG flow is amended by an adequate optimisation. The
    results are also evaluated in the light of stability and minimum
    sensitivity considerations. Extensions to higher order and further
    implications are emphasized.}
\end{minipage}
\end{center}

\newpage
\noindent 
{\bf 1. Introduction}\\[-1ex]

It is an experimental fact that physical correlation lengths are diverging in
the vicinity of a critical point like a second order phase transition. The
absence of dimensionful length scales implies scale invariance of physical
correlation functions. Then, properties of physical systems close to a scaling
regime or a critical point are characterised by universal
exponents and scaling laws \cite{Zinn-Justin:1989mi}. \step

In quantum field theory, functional methods like the renormalisation group are
important tools in the study of strongly coupled systems and critical
phenomena.  The Wilsonian renormalisation group is based on the successive
integrating-out of momentum degrees of freedom from a path integral
representation of the theory, thereby interpolating between a given classical
theory and the full quantum effective action
\cite{Polchinski,continuum,Ellwanger:1994mw,Morris:1994qb} (for reviews, see
\cite{reviews}).  The strength of these methods is their flexibility when it
comes to approximations, in particular for theories with strong couplings or
large correlation lengths. Furthermore, powerful optimisation criteria are
available to increase the reliability within given truncations
\cite{Litim:2000ci,Litim:2001up,Litim:2001fd}.  Wilsonian flows play an
important role in the study of universal scaling phenomena in gauge theories
and gravity \cite{Pawlowski:2003hq,gauge,gravity}.\step

In this Letter, we compare different implementations of a Wilsonian cut-off,
the Exact Renormalisation Group based on an infrared momentum cut-off for the
full effective action $\Gamma$, and the Polchinski renormalisation group based
on an ultra\-violet momentum cut-off for the action $S$. Both approaches
correspond to exact flows, meaning that the endpoint of the fully integrated
flow is given by the physical theory. This equivalence, in general, is lost in
given truncations due to the qualitative differences in these approaches. We
analyse the structural differences both in the full flows and within a
derivative expansion.  We also establish the remarkable result that the
universal information encoded within the Polchinski renormalisation group to
leading order in a derivative expansion is equivalent to the Exact
Renormalisation Group, if the latter is amended by an adequate
optimisation. Extensions to higher order and further implications of this
result are equally discussed.\\[3ex]

\noindent
{\bf 2. Generalities}\\[-1ex]

Wilsonian flows are based on integrating-out momentum modes from a path
integral representation of quantum field theory. A Wilsonian flow connects a
short-distance effective action - typically the classical action - with the
full physical theory. Given that the main physical information is contained in
the integrated flow, the key properties of the different implementations
proposed in the literature deserve a more detailed study.  \step

The Exact Renormalisation Group is based on a cutoff term $\Delta
S_k=\s012\int\, \phi\, R\,\phi$, added to the Schwinger functional. The
operator $R(q)$ introduces a momentum cutoff at momentum scale $q^2\approx
k^2$. This induces a $k$-dependence on the level of the effective action.  In
its modern form, the flow for an effective action $\Gamma_k$ for bosonic
fields $\phi$ is given by the simple one-loop expression
\cite{continuum,Ellwanger:1994mw,Morris:1994qb}
\beq \label{ERG}
\partial_t\Gamma_k[\phi] =
\frac{1}{2}\Tr\,\frac{1}{{\Gamma^{(2)}_k}+ R}\, \partial_t  R
\eeq
Here, $t\equiv\ln k$ is the logarithmic scale parameter, and the trace denotes
a loop integration and a sum over fields and indices, and
$\Gamma_k^{(2)}[\phi](p,q)\equiv \delta^2\Gamma_k/\delta\phi(p)\delta\phi(q)$.
The regulator $R$ can be chosen freely, though within a few restrictions which
ensure that the flow is well-defined, thereby interpolating between an initial
action in the ultra\-violet (UV) and the full quantum effective action in the
infrared (IR) \cite{reviews}. In momentum space, the flow \eq{ERG} receives
its main contributions for momenta in the vicinity of $q^2\approx k^2$,
because large momentum modes are suppressed by $\partial_t R$, and small
momentum modes are suppressed because $R$ is an IR cutoff. \step

The flow \eq{ERG} depends on fields and couplings only through the full
propagator. More generally, any exact RG flow for $\Gamma_k$ with a one-loop
structure always depends linearly on the full propagator, and hence on the
inverse of $\Gamma^{(2)}_k$ \cite{Litim:2002xm}. The linear dependence on the
full propagator implies that an exact flow for $\Gamma_k$ is small whenever
the full propagator is quantitatively small in the momentum regime where
$\partial_t R$ is peaked, $e.g.$~for large fields or strong coupling.  Hence,
the Wilsonian flow \eq{ERG} is essentially local both in field- and
momentum-space. Owing to this structure, the flow is amiable to truncations
local in the fields (vertex functions) in the range where \eq{ERG} is
non-trivial.  An integration of the flow in given truncations requires a good
control over the full propagator.  The stability of \eq{ERG} -- in the regions
of momentum- and field-space where it is non-trivial -- is controlled by the
cutoff propagator, and hence by the momentum cutoff $R$ \cite{Litim:2000ci}.
Therefore, the convergence of truncated flows can be increased by appropriate
choices of $R$. Efficient optimisation criteria based on the flow are
available \cite{Litim:2000ci,Litim:2001up,Litim:2001fd}.\step

A different version of an exact renormalisation group has been introduced by
Polchinski \cite{Polchinski}, and is based on an ultraviolet regulator
$K(q^2/\Lambda^2)$ for propagators in the path integral, where $\Lambda$
denotes the ultraviolet scale parameter.  The Polchinski flow equation for the
Wilsonian action $S_\Lambda[\phi]$ for a scalar field theory is given by
\cite{Ball:1995ji}
\beq\label{PRG}
\partial_t S_\Lambda[\phi]=
\012 \Tr\, \, \partial_t P_\Lambda(q)\,
\left[S_\Lambda^{(1)}(q)\, S_\Lambda^{(1)}(-q)
-S_\Lambda^{(2)}(q,-q)
-2 (P_\Lambda(q))^{-1} \phi(q)\,S_\Lambda^{(1)}(q)\right]\,,
\eeq
where $S^{(1)}[\phi](q)\equiv \delta S[\phi]/\delta\phi(q)$,
$P_\Lambda(q)=K(q^2/\Lambda^2)/q^2$ and $t\equiv \ln \Lambda$. The cutoff
function $K$ is chosen such that high momentum modes are suppressed,
$K(q^2/\Lambda^2\to\infty)\to 0$. The scale dependence introduced via $K$
induces the scale dependence of $S_\Lambda[\phi]$ on $\Lambda$. By
construction, the flow equation \eq{PRG} is exact in the same sense as the ERG
flow \eq{ERG}. \step

The Polchinski flow \eq{PRG} depends on the fields via functional derivatives
of $S$ -- linearly through the operators $S^{(2)}$ and $P^{-1}_\Lambda
\,\phi\,S^{(1)}$, and non-linearly through $S^{(1)}\, S^{(1)}$. Each of these
terms can grow large for large fields and/or large couplings, even in the
momentum regime where $\partial_t P_\Lambda$ is non-vanishing. In general,
they do not cancel amongst each other. Hence, the right-hand side of \eq{PRG}
remains non-trivial in large parts of field space. This makes it more
difficult, within given truncations, to identify stable flows or regions in
field space where the flow remains small. On the other hand, the polynomial
non-linearities in \eq{PRG} are simpler than those in \eq{ERG} as they do not
involve an inverse of $S^{(n)}$, also leading to a smaller number of different
terms in the flow for a $n$-point function $\partial_t S^{(n)}$ for
sufficiently large $n$.  These aspects have been considered as a major benefit
of the Polchinski flow, $e.g.$~\cite{Ball:1995ji}. Below,  we shall see 
that it is precisely the non-linear dependence of \eq{ERG} on the inverse of
$\Gamma_k^{(2)}$ which implies stability of the flow, whereas the polynomial
non-linearities in \eq{PRG} are responsible for less stable solutions.\step

Flow equations based on a proper-time regularisation have received
considerable interest recently
\cite{Liao:1994fp,Bohr:2000gp,Litim:2001hk,Litim:2002hj}.  They derive from a
proper-time regularisation of the effective action valid to one-loop order
\cite{Liao:1994fp}. Any proper-time flow at momentum scale $k$ can be
represented in the basis of
\beq \label{PTRG}
\partial_t\Gamma_k[\phi] =
\frac{1}{2}\Tr \left( \frac{k^2}{k^2 +\Gamma_k^{(2)}/m}\right)^m\,,
\eeq
or linear combinations thereof, where the parameter $m\ge 1$ characterises the
momentum cutoff \cite{Litim:2001hk,Litim:2002xm}.  Path integral derivations
of proper-time flows have been worked out in
\cite{Litim:2001hk,Litim:2002xm,Litim:2002hj}. In their simplest form, they
make use of background fields, where plain momenta $q^2$ in the cutoff is
replaced by $\Gamma_k^{(2)}[\phi]$ evaluated at some background field
\cite{Litim:2002xm}. Therefore, an exact proper-time flow contains additional
flow terms proportional to $\partial_t \Gamma^{(2)}_k$ on its right-hand side
\cite{Litim:2002hj}. Implicit to this approach is that differences between
fluctuation and background field are neglected.  The proportionality of
\eq{PTRG} to (powers of) the full propagator is responsible for an increased
suppression of the flow at large fields or couplings with increasing $m$. In
this respect, proper-time flows are similar to ERG flows and, consequently,
allow for analogous optimisations \cite{Litim:2001up,Litim:2001dt}.  Given the
established link between exact proper time flows and ERG flows, it suffices,
in the remainder, to elaborate on the differences between \eq{ERG} and
\eq{PRG}.\\[3ex]

\noindent
{\bf 3. Derivative expansion}\\[-1ex]

Both \eq{ERG} and \eq{PRG} are exact flows. In consequence, the physical
content of the fully integrated flows should be identical. This equivalence,
in general, cannot be maintained within specific approximations, unavoidable
as soon as either method is applied to a non-trivial physical
problem. In order to highlight the similarities and differences between
\eq{ERG} and \eq{PRG}, we study both flows for an $O(N)$ symmetric real scalar
field $\phi^a$, $a=1\cdots N$, in $d$ dimensions within a derivative expansion
\cite{Golner:1986}, which is the most commonly used expansion scheme for
critical phenomena (see also \cite{Pawlowski:2003hq,Blaizot:2004}). A priori,
little is known about its convergence as there is no small expansion parameter
associated to it \cite{Litim:2001dt}. The ERG flow for
$\Gamma_k$, and the Polchinski flow for $S_\Lambda$ are linked by a Legendre
transform and additional momentum-dependent field rescalings.  This implies
that derivative expansions for the ERG and the Polchinski RG are inequivalent.
To leading order in the derivative expansion, an Ansatz for the effective
action $\Gamma_k$ contains a standard kinetic term and the effective potential
$U_k$,
\beq\label{AnsatzGamma}
\Gamma_k=\int d^dx \left(U_k(\bar\rho) 
             + \012 \partial_\mu \phi^a\partial_\mu \phi_a
\right)\ ,
\eeq
where $\bar\rho=\s012\phi^a\phi_a$.  Introducing dimensionless variables
$\rho=\bar\rho k^{2-d}$, $U(\bar\rho)=u(\rho)\,k^d$, and using \eq{ERG}, the
flow equation for the effective potential is
\beq\label{ERG-Flow}
\partial_t u+du-(d-2) \rho u'
=(N-1)\, \ell(\omega_1)+\, \ell(\omega_2)\,
\eeq
with $\omega_1=u'$ and $\omega_2=u'+2\rho u''$. The function $\ell(\omega)$
encodes the non-trivial flow, and reads
\beq\label{Id}
\ell(\omega)
=v_d\int^\infty_0dy\, y^{d/2}\, \0{\partial_t r(y)}{y(1+r)+\omega}\,
\eeq
with $y\equiv q^2/k^2$, $r(y)=R(q^2)/q^2$, $\partial_t r(y) = -2 y r'(y)$, and
$v^{-1}_d=2^{d+1}\pi^{d/2}\Gamma(d/2)$. The flow \eq{ERG-Flow} is a second
order non-linear partial differential equation. All non-trivial information
regarding the renormalisation flow and the regularisation scheme are encoded
in the function \eq{Id}. The momentum integration is peaked and regularised
for large momenta due to the cutoff term $\partial_t r(y)$, and for small
momenta due to $r(y)$ in the numerator.\step

All terms on the left-hand side of \eq{ERG-Flow} are cutoff independent, and
display the intrinsic scaling of the variables that have been chosen for the
parametrisation of the flow. By making use of rescaling in the fields and in
the effective potential, the numerical factor $\sim v_d$ can be
removed. Rescalings of the fields and the infrared scale parameter cannot
remove the explicit cutoff dependence in \eq{Id}. The $R$-dependence of the
flow \eq{ERG-Flow} can be characterised by appropriate moments of $R$
\cite{Litim:2000ci}.\step

In \cite{Litim:2000ci,Litim:2001up,Litim:2001fd,Litim:2001dt}, ideas have been
put forward to increase the stability and physical content of truncated RG
flows by choosing `optimised' regulators $R$. The main observation is that
the infrared cutoff $R$, in addition to regularising the flow, also controls
its convergence and stability properties. This fact entails that specific
cutoffs lead to improved results already at a fixed order in a systematic
expansion. An optimisation then corresponds to identifying the RG flows with
best stability properties.  As an example, consider the flow \eq{ERG} in an
expansion in vertex functions about vanishing field. The truncated propagator
is $G_k(q^2)=(q^2+R)^{-1}$. Its contribution to the flow is largest in the
momentum range where $G_k(q^2)$ is maximal. As a function of momenta, the
cutoff propagator in units of $k$ achieves the maximum
$C^{-1}(R)=\max_{q^2/k^2}\ \left[G_k(q^2) \, k^2\right]$. Consequently, the
flow \eq{ERG} displays an increased stability if the maximal propagator
contributions $\sim C^{-1}(R)$ remain as small as possible, $e.g.$~for
regulators $R$ for which the 'gap parameter' $C(R)$ becomes maximal,
\beq\label{Opt}
\max_R\ C(R)\,.
\eeq
This condition requires an appropriate normalisation for the cutoff.
Hence, \eq{Opt} states that the gap $C(R)$
should be maximal with respect to the cutoff function $R$.  To leading order,
the gap criterion, and its solution, is independent of the specific theory
studied. Note that \eq{Opt} is a very mild condition: it only fixes
one parameter in $R$ out of infinitely many. This implies that the subspace of
optimised cutoffs is still infinite dimensional.\step

Optimised flows have a number of interesting properties: their radius of
convergence for amplitude expansions is increased \cite{Litim:2000ci}, they
factorise thermal and quantum fluctuations in the flow \cite{Litim:2001up},
they entail a minimum sensitivity condition \cite{Litim:2001fd}, they improve
the derivative expansion \cite{Litim:2001dt}, they lead to a fast decoupling
of heavy modes, and they lead to an improved approach to convexity for
theories with spontaneous symmetry breaking.  An important optimised cutoff is
given by \cite{Litim:2001up}
\beq\label{Ropt} 
R_{\rm opt}(q^2) = (k^2-q^2)\,\theta (k^2-q^2)\,. 
\eeq
In momentum space, the cutoff \eq{Ropt} is distinguished because it solves
\eq{Opt} in the entire domain $q^2<k^2$, and not only at the minimum of the
inverse cutoff propagator. Hence, \eq{Ropt} implements the gap criterion in a
global manner. In the space of all optimised cutoff propagators, the cutoff
\eq{Ropt} corresponds to the convex hull of optimised inverse cutoff
propagators (cf.~Fig.~1 in \cite{Litim:2001up}), reflecting the extremal
property of \eq{Ropt}.  In more physical terms, the cutoff \eq{Ropt} leaves
the propagation of large momentum modes $q^2>k^2$ unchanged, $q^2+R\approx
q^2$. In turn, the propagation of infrared modes with $q^2<k^2$ is cut off
leading to an effective mass term $q^2+R\approx k^2$.  When expressed in terms
of \eq{Ropt}, the flow \eq{ERG-Flow} becomes
\beq\label{ERG-FlowOpt}
\partial_t u +d u -(d-2) \rho u'
=                \0{N-1}{1+u'} 
                +\0{1}{1+u'+2\rho u''}.
\eeq
The numerical factor $4v_d/d$ has been absorbed into the potential and
the fields by an appropriate rescaling.\step

Now we turn to the Polchinski flow.  We use an Ansatz for $S_\Lambda[\phi]$
analogous to \eq{AnsatzGamma}.  For comparison with \eq{ERG-Flow}, we
introduce the dimensionless effective potential $u(\rho)=U_\Lambda/\Lambda^d$
and the dimensionless field variable $\rho=\s012 \phi^a\phi_a \Lambda^{2-d}$
Then, for $N\neq 0$, the Polchinski flow reads
\cite{Ball:1995ji,Comellas:1997tf,Comellas:1998ep}
\beq\label{PRG-Flow}
\partial_t u -d u +(d-2) \rho u'
=
u'+\s02N \rho u'' -2\rho(u')^2 \,.
\eeq
We also have performed a finite renormalisation of the fields and the
potential.  The crucial observation at this point is that the flow equation
\eq{PRG-Flow} is cutoff independent. The main differences and similarities
between the ERG flow \eq{ERG-Flow}, \eq{ERG-FlowOpt} and the Polchinski flow
\eq{PRG-Flow} are summarised as follows:\\[-2ex]

$(i)$ {\it Non-linearities---}The essential nonlinearities in the
 flows \eq{ERG-Flow} and \eq{PRG-Flow} are very
different. For the Polchinski flow, they reduce to a quadratic term $\sim
\rho\cdot (u')^2$. For the ERG flow, the non-linearities appear solely in a
denominator, a direct consequence of the structural form of \eq{ERG}. When
expanded in powers of the fields, the flow contains all and
arbitrarily high powers of $u'$, $2 \rho u''$ and products thereof. \\[-2ex]

$(ii)$ {\it Stability---}The structure of the non-linearities influence the
stability properties of the flows. In the strong coupling domain or at large
fields, the non-trivial part of the ERG flow (the right-hand side of
\eq{ERG-Flow}) is small, effectively suppressed by powers of the
propagator. The non-linearities of the Polchinski flow (the right-hand side of
\eq{PRG-Flow}) are unbounded for large fields and couplings.\\[-2ex]

$(iii)$ {\it Cutoff independence---}The terms on the right-hand sides of
\eq{ERG-Flow} and \eq{PRG-Flow} originate from the non-trivial flow
of the potential and contain the essential non-linearities. Their structure
constitutes the main qualitative difference between the two flows. 
The ERG flow \eq{ERG-Flow} depends on infinitely many moments of the regulator
$R$. This can be seen explicitly by expanding the flow in powers of $\omega_1$
and $\omega_2$. On a scaling solution, the Polchinski flow \eq{PRG-Flow} is
fully scheme independent. This is in marked contrast to the ERG flow, where
the regulator dependence cannot be removed by rescalings of the fields. A weak
scheme dependence may persist even after an integration of the truncated
flow.\\[3ex]

\begin{figure}
\begin{center}
\unitlength0.0006\hsize
\begin{picture}(1000,815)
\put(320,400){ ${\nu_{\rm opt} = 0.649562\cdots}$}
\put(320,350){ ${\nu_{\rm phys} \approx 0.63}$}
\put(870,440){ ${\nu_{\rm min}}$}
\put(870,480){ ${\nu_{\rm sharp}}$}
\put(870,670){ ${\nu_{\rm max}}$}
\put(440,230){ \large $R$}
\put(40,780){ \Large $\nu$}
\hskip.04\hsize
\epsfig{file=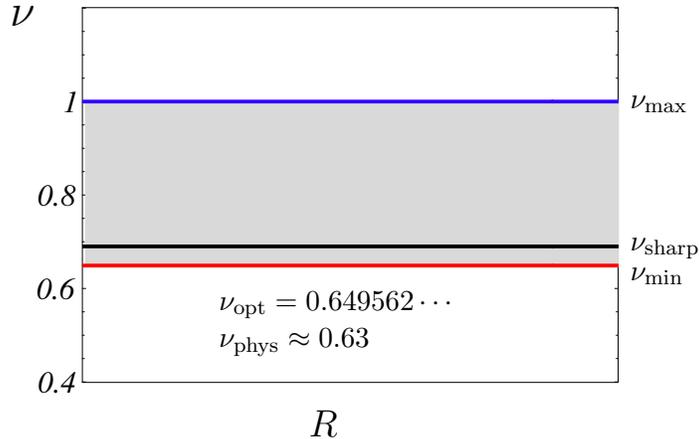,width=800\unitlength}
\end{picture}
\vskip-1.8cm 
\caption{Critical exponent $\nu$ for various $R$, $N=1$. The shaded region
  contains about $10^3$ data points for various classes of cutoffs. We have
  indicated the sharp cutoff result (black line), the numerical result
  $\nu_{\rm max}$ at the upper boundary (blue line) and at the lower boundary
  $\nu_{\rm min}=\nu_{\rm opt}$ (red line). The latter coincides with the
  result from the optimised ERG flow \eq{ERG-FlowOpt} and the Polchinski flow
  \eq{PRG-Flow}. The physical value is also indicated (see text).}
\end{center}
\end{figure}

\noindent
{\bf 4. Universality}\\[-1ex]

Next, we switch to $d=3$ dimensions and compare the Polchinski and ERG flows
on a quantitative level in the vicinity of a scaling solution, the
Wilson-Fisher fixed point.  The numerical values for universal critical
exponents from Polchinski flows, due to their scheme independence to leading
order in a derivative expansion, can be seen as a benchmark test for any other
approach in the same approximation. For ERG flows, this check is non-trivial
since \eq{ERG-Flow} depends on the cutoff.\step

Universal critical exponents $\nu$ and subleading corrections-to-scaling
exponents have been deduced in the literature from the non-trivial scaling
solution $\partial_t u'_*=0$ as eigenvalues $\lambda$ of small perturbations
$\delta u'$ around the fixed point, $\partial_t(u'_*+\delta u')=\partial_t\
\delta u'=\lambda\ \delta u'$ ($\nu=-1/\lambda_0$, where $\lambda_0$ is the
single negative eigenvalue).  The flow $\partial_t\ \delta u'$ and the
eigenperturbations $\delta u'$ depend on the scaling solution $u'_*$. \step

We begin with the ERG flow \eq{ERG-Flow}. The maximal ranges of attainable
values for the critical exponents $\nu(R)$, for all $N$, has been discussed in
\cite{Litim:2002cf}. The main result is depicted in Fig.~1 for $N=1$, which
contains $\sim 10^3$ data points for $\nu(R)$, based on qualitatively
different classes of cutoff functions including exponential, compact and
algebraic ones, combinations thereof, discontinuous cutoffs and cutoffs with
sliding scales. The main result represented by Fig.~1 is that the range of
values for $\nu(R)$ is bounded both from above and from below. The numerical
value at the upper bound
\beq\label{max}
\nu_{\rm max}=\max_{R}\ \nu(R)\,,
\eeq
corresponds to the large-$N$ limit $\nu =1$.  The upper boundary is achieved
for a Callan-Symanzik type flow with mass-like regulator $R\to k^2$. In this
limit, the corresponding flow ceases to be a Wilsonian flow in the strict
sense because the momentum integration in \eq{ERG} is no longer cut-off in the
ultra\-violet limit. In the light of the optimisation, these flows have poor
convergence and stability properties, and do not represent solutions to
\eq{Opt}. This behaviour is equally reflected in the increasingly poor
numerical convergence of solutions to \eq{ERG-Flow} for cutoffs with $\nu$ in
the vicinity of \eq{max}. Conversely, in the vicinity of the lower boundary
\beq\label{nu-Opt}
\nu_{\rm min}=\min_{R}\ \nu(R)\,,
\eeq
the flows have good convergence and stability behaviour. Results $\nu(R)$ from
generic optimised flows are typically less than 1\% away from the lower
boundary \eq{nu-Opt}. Hence, the numerical value $\nu_{\rm min}$ and all
regularisations leading to it, are distinguished. Most interestingly, the
minimum \eq{nu-Opt} is achieved for the cutoff \eq{Ropt},
\beq\label{min}
\nu_{\rm opt}\stackrel{!}{=}\nu_{\rm min}\,.
\eeq
Therefore, the value \eq{min} has maximal reliability in the present
 truncation and is taken as the physical prediction to this order.\step

Fig.~1 can also be interpreted in the light of the principle of minimum
sensitivity (PMS) \cite{Stevenson:1981vj}. We emphasize that the latter is
only applicable because $\nu(R)$ is globally bounded. Then, the PMS condition
corresponds to the choice of cutoffs $R_{\rm PMS}$, for which $\delta
\nu(R)/\delta R$ vanishes. For one-parameter families of cutoffs $R(b)$
parametrised by $b$, the PMS condition often has several solutions for $b_{\rm
PMS}$ and $\nu_{\rm PMS}$, $e.g.$~\cite{Liao:1999sh,Litim:2001fd}. Requiring
that $\nu(R)$ is `globally' extremal with respect to the regularisation
identifies both boundaries \eq{max} and \eq{min} as solutions of a global
minimum sensitivity condition. Here, `global' refers to the fact that the
extrema in Fig.~1 are achieved within the entire space of regulators $R$, and
not just `locally' for some $n$-parameter subclasses thereof (see
\cite{Litim:2002cf}). Hence, the PMS condition by itself, neither locally nor
globally, is sufficient to provide a unique physical prediction for $\nu$. In
turn, the optimisation condition singles out a unique prediction: locally, for
one-parameter families of cutoffs, it leads to values for $\nu$ in the
vicinity of \eq{min} \cite{Litim:2001dt}; globally, the value \eq{min} is
singled out straightaway. Hence, the optimisation which has let to the choice
\eq{Ropt} is equivalent to a global extremisation of $\nu(R)$. This provides
an explicit example for the more general result of \cite{Litim:2001fd}, which
states that the optimisation entails a minimum sensitivity condition, while
the converse, in general, is not true.\step

\begin{table}
\begin{tabular}{c|lc}
$\quad N\quad$&
$\quad{\rm Polchinski}\quad$&
$\quad\quad R_{\rm opt}\quad\quad$\\[.5ex] \hline
$0$&
${}\quad$
${}\quad$
---&
$\ 0.592083^{a}\ {}$\\
$1$&
${}\quad$
$\ 0.64956^{c,e} {}$&
$\ 0.649562^{a}\ {}$\\
$2$&
${}\quad$
$\ 0.7082^{d}\ \  {}$&
$\ 0.708211^{a}\ {}$\\
$3$&
${}\quad$
$\ 0.7611^{d}\ {}$&
$\ 0.761123^{a}\ {}$\\
$4$&
${}\quad$
$\ 0.8043^{d}\ {}$&
$\ 0.804348^{a}\ {}$\\[-1ex]
\end{tabular}
\caption{Critical exponent $\nu$ (see text).}
\end{table}

We continue with a comparison of all critical exponents that have been
published to date within both the optimised ERG flow \eq{ERG-FlowOpt} and the
Polchinski flow \eq{PRG-Flow}. These are: the critical exponent $\nu$ (Table
I), the smallest correction-to-scaling exponent $\omega$ (Table II) and the
asymmetric corrections-to-scaling exponent $\omega_5$ (Table III).  Based on
the optimised ERG flow \eq{ERG-FlowOpt}, the exponents $\nu$ and $\omega$ have
been given for all $N$ in \cite{Litim:2002cf} with up to six significant
figures $(a)$. The exponent $\omega_5$ has been computed in
\cite{Litim:2003kf} for $N=1$ $(b)$.  Cutoff independence of the Polchinski
flow implies unique results for universal eigenvalues. The critical exponents
$\nu$ and $\omega$ have been computed with up to four significant digits in
\cite{Ball:1995ji} for $N=1$ $(c)$, and in \cite{Comellas:1997tf} for
$N=1\cdots 4$ $(d)$. Results up to five digits for the exponents $\nu, \omega$
and $\omega_5$ have recently been stated in \cite{Bervillier:2004mf} for $N=1$
$(e)$. Except for $N=\infty$, there are no published results based on the
Polchinski flow for $N>4$. However, it has recently been indicated
\cite{Bervillier:2005za} that the results from Polchinski flow and optimised
ERG flow also agree for $N>4$. In the large $N$ limit, the
spread of $\nu(R)$ with $R$ is absent, and the results for critical exponents
becomes unique, $\nu(R)=1$, and $\omega_n(R)=2n-1$, $n=1,2,\cdots$ in
agreement with the corresponding limit of the Polchinski flow
\cite{D'Attanasio:1997ej}. \step

Hence, it is most remarkable that all universal critical exponents computed
either from the Polchinski flow or from the optimised ERG flow, agree to all
significant figures, for the leading and subleading critical exponents, and
for different universality classes! This high degree of coincidence leads to
the important conjecture that the universal content of the partial
differential equations \eq{ERG-FlowOpt} and \eq{PRG-Flow} is equivalent
\footnote{After publication, an explicit map between \eq{ERG-FlowOpt} and
\eq{PRG-Flow} has been worked out by T.~R.~Morris in
hep-th/0503161.
}.
\\[3ex]

\begin{table}
\begin{tabular}{c|lc}
$\quad N\quad$&
$\quad {\rm Polchinski}\quad$&
$\quad\quad R_{\rm opt}\quad\quad$\\[.5ex] \hline
$0$&
${}\quad$
${}\quad$
---&
$\ 0.65788^{a}\ \ \ {}$\\
$1$&
${}\quad$
$\ 0.65574^{c,e}\ {}$&
$\ 0.655746^{a}\ {}$\\
$2$&
${}\quad$
$\ 0.6712^{d}\ {}$&
$\ 0.671221^{a}\ {}$\\
$3$&
${}\quad$
$\ 0.6998^{d}\ {}$&
$\ 0.699837^{a}\ {}$\\
$4$&
${}\quad$
$\ 0.7338^{d}\ {}$&
$\ 0.733753^{a}\ {}$
\end{tabular}
\caption{Subleading correction-to-scaling exponent $\omega$.}
\end{table}

\noindent
{\bf 5. Stability}\\[-1ex]

Next, we analyse the locality and stability structure of flows and show that
the non-universal properties of RG flows in the vicinity of a scaling solution
are vastly different. Consider the fixed point solution itself, which is
non-universal and not measurable in any experiment.  Using \eq{PRG-Flow}, the
nontrivial scaling solution with $u'_\star \neq$ const.~and $\partial_t
u'_\star=0$ of the Polchinski RG obeys the differential equation
\beq\label{FlowPRGupScal}
\nonumber
2 u_\star' - (d-2)\rho u_\star''
   =2(u_\star')^2 - \s02N \rho u_\star''' 
    -(1+\s02N-4\rho u_\star')u_\star''\,.
\eeq
For large fields, the scaling potential behaves as
\beq\label{solutionPRGExpansionLarge}
u_\star(\rho) \propto\rho +{\rm subleading}\,.
\eeq
An analytical solution for $N=\infty$ has been given in
\cite{Kubyshin:2001gz}. On the level of the RG flow, this behaviour stems from
a cancellation between the canonical scaling of the potential and its
non-linear renormalisation. From \eq{solutionPRGExpansionLarge}, we conclude
that the
non-trivial quantum contributions to the 
Polchinski flow diverges like
\beq\label{PRG-largefield}
\partial_t u -d u +(d-2) \rho u'
\propto \rho+{\rm subleading}
\eeq
for large fields $\rho$ close to a scaling solution.  Within the optimised
ERG, the non-trivial scaling solution, using \eq{ERG-FlowOpt}, obeys
\beq
\nonumber
\label{FlowPotential3} 
2 u'_\star - (d-2)\rho u''_\star =
-(N-1) \0{u''_\star}{(1+u'_\star)^2} 
-\0{3u''_\star+2\rho u'''_\star}{(1+u'_\star+2\rho u''_\star)^2} \,.
\eeq
Analytical solutions for the limit $N=\infty$ have been given in
\cite{Litim:1995ex}. For $N\neq\infty$, and in the vicinity of $\rho=0$, the
scaling solution can be obtained analytically as a Taylor expansion in the
field \cite{Litim:2002cf}. In the limit of large fields $\rho\gg 1$, we find
\beq\label{solutionExpansionLarge}
u_\star(\rho)\propto \rho^{1+\alpha} +{\rm subleading}
\eeq
for arbitrary regulator, where $\alpha=2/(d-2)$ is positive for $d>2$. Here,
the large-field behaviour is solely due to the canonical scaling dimension of
the fields. From \eq{FlowPotential3}, and for large fields $\rho \gg 1$, we
conclude that the non-linear part of the ERG flow for the potential behaves as
\beq\label{ERG-largefield}
\partial_t u + du -(d-2) \rho u'
\propto \rho^{-\alpha}+{\rm subleading}\,.
\eeq
Hence, the right-hand side of \eq{ERG-largefield} is suppressed for all $d>2$.
\step

\begin{table}
\begin{tabular}{c|cc}
$\quad N\quad$&
$\quad {\rm Polchinski}\quad$&
$\quad\quad R_{\rm opt}\quad\quad$\\[.5ex] \hline
$1$&
$\ 1.8867^{e}\ {}$&
$\ 1.8867^{b}\ {}$
\end{tabular}
\caption{Asymmetric correction-to-scaling exponent $\omega_5$.}
\end{table}

More generally, the result \eq{ERG-largefield} holds for generic ERG flows
where the right-hand side of \eq{ERG-Flow} decays $\propto 1/\omega$ for large
$\omega$. The power-law behaviour is altered for cutoffs which effectively
introduce non-localities due to their momentum structure, $e.g.$~the mass-like
cutoff (no large momentum decay), the sharp cutoff, or cutoffs with an
algebraic large-momentum decay like the quartic cutoff $R\sim k^4/q^2$, {\it
e.g.}  \cite{Morris:1997xj}. They lead, respectively, to \eq{ERG-largefield}
with a large-field behaviour $\propto \rho$, $\propto \ln\rho$ and, in three
dimensions, $\propto \rho^{-3/2}$.  The different power law exponents
$\alpha(R)$ as a function of the cutoff are displayed in Fig~2 for three
dimensions.  The minimum
\beq
\alpha_{\rm min}=\min_R\ \alpha(R)
\eeq
is achieved for Callan-Symanzik type flows and the Polchinski flow,
$\alpha_{\rm min}=-1<0$.  A negative $\alpha$ also indicates that an
additional renormalisation of the flow is necessary due to an insufficiency in
the integrating-out of momentum modes. This is well-known for Callan-Symanzik
type flows \cite{reviews}. We stress, however, that the set of flows with
negative $\alpha$ is of measure zero; generic ERG flows have positive
$\alpha$. The sharp cutoff marks the boundary between ERG flows with
insignificant ($\alpha>0$) and significant ($\alpha<0$) contributions for
large fields, and hence the boundary between flows which are essentially
local, respectively non-local, in the fields. The maximum
\beq
\alpha_{\rm max}=\max_R\ \alpha(R)
\eeq
is achieved for generic ERG flows including optimised ones, $\alpha_{\rm
max}=2/(d-2)>0$. Note that the few 'non-local' ERG flows with $\alpha$ in the
range $[\alpha_{\rm min},\alpha_{\rm sharp}]$ lead to critical exponents $\nu$
in Fig.~1 in the range $[\nu_{\rm max},\nu_{\rm sharp}]$, whereas all 'local'
ERG flows with $\alpha$ within $[\alpha_{\rm sharp},\alpha_{\rm max}]$ -- the
overwhelming majority of all ERG flows -- lead to values within the narrow
window $[\nu_{\rm sharp},\nu_{\rm min}]$ \cite{Litim:2002cf}. We conclude that
flows with underlying non-localities have the tendency to deviate strongly
from the physical theory and display a strong cutoff dependence, while local
flows display only a weak cutoff dependence, thereby remaining close to the
physical theory. In the Ising universality class, the exponent $\nu$ from
non-local (local) flows deviates between 10-50\% (3-10\%) from the physical
value.  This quantifies the link between the locality structure of the flow,
its stability, and its vicinity to the physical theory. \step

Summarising, unlike the universal parts the non-universal scaling solutions
derived from ERG or Polchinski flows are vastly different. This result also
extends to the non-universal eigenperturbations at criticality.  The
non-trivial quantum corrections to the flow at large fields are strongly
suppressed for generic ERG fows \eq{ERG-largefield}, while they remain large
in the Polchinski case \eq{PRG-largefield}. This is a direct consequence of
the structural differences in the basic flows \eq{ERG} and \eq{PRG}.  Despite
of having the same universal content, in the light of Fig.~2 the optimised
flow \eq{ERG-FlowOpt} and the Polchinski flow \eq{PRG-Flow} have maximally
distinct locality structures. Note that good locality properties of flows are
at the root for stable numerical integrations, and the quantitative smallness
of quantum corrections in large domains of field space improves the
convergence of RG flows.  Based on the result that ERG flows with increasing
non-localities show an increasingly strong cutoff dependence, we expect that
Polchinski flows display a similar behaviour as soon as the leading-order
degeneracy with respect to the cutoff is lifted by higher order operators in
extended truncation.  \\[3ex]

\begin{figure}
\begin{center}
\unitlength0.0006\hsize
\begin{picture}(1000,815)
\put(900,400){$\alpha_{\rm min}$}
\put(900,500){$\alpha_{\rm sharp}$}
\put(900,710){$\alpha_{\rm max}$}
\put(450,230){ \large $R$}
\put(40,800){\large $\alpha$}
\hskip.04\hsize
\epsfig{file=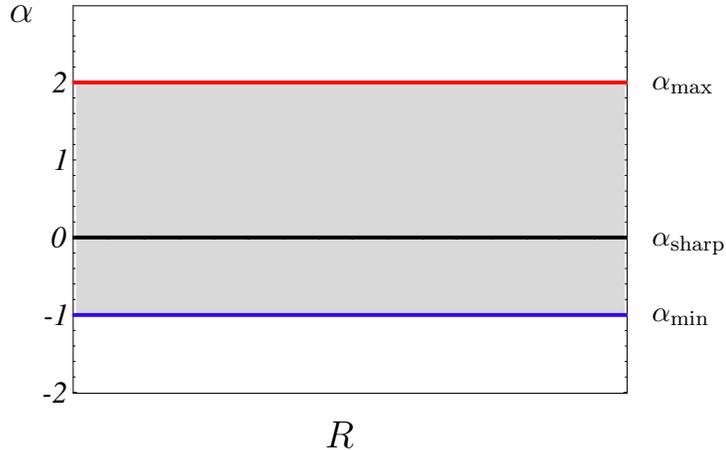,width=800\unitlength}
\end{picture}
\vskip-1.8cm 
\caption{ Large-field behaviour of Wilsonian flows close to a critical point,
$d=3$. The shaded region indicates the range of values for $\alpha(R)$ for
various $R$. The sharp cutoff (black line) marks the boundary between flows
with significant ($\alpha<0$) and insignificant ($\alpha>0$) contributions at
large fields. Callan-Symanzik type flows and the Polchinski flow have
$\alpha=\alpha_{\rm min}$ (blue line), generic ERG flows have
$\alpha=\alpha_{\rm max}$ (red line). Flows with an increased locality
structure lead to improved results (see text).  }
\end{center}
\end{figure}

\noindent
{\bf 6. Extensions}\\[-1ex]

A global analysis of critical exponents in the full space of cutoffs has only
been performed to leading order in the derivative expansion, which is the most
important order as higher order effects should be suppressed proportional to
the anomalous dimension $\eta$ of the order of a few percent. Still, it is
useful to briefly review the results achieved so far beyond leading order in
the light of the preceeding discussion.\step

Within Polchinski flows, the leading order scheme independence is lost as soon
as higher order derivative operators are taken into account
\cite{Ball:1995ji}. To order ${\cal O}(\partial^{n})$, the flow depends
explicitely on $n$ scheme-dependent parameters which cannot be removed by
further rescalings.  To order ${\cal O}(\partial^{2})$, a part of the cutoff
dependence has been probed for different projections on the anomalous
dimension, {\it
i.e.}~\cite{Ball:1995ji,Comellas:1998ep,Bervillier:2004mf,Bervillier:2005za}.
All published results for $\nu$ and $\eta$ to order ${\cal O}(\partial^{2})$
have in common that the spurious dependence on remaining unphysical cutoff
parameters is monotonous, without displaying local extrema.  The range of
numerical results includes the physical values. Unfortunately, none of the
truncations admits a minimum sensitivity condition, and further conditions
have to be invoked to remove the scheme dependence.  From a structural point
of view, the comparatively strong cutoff dependence beyond leading order is
not unexpected and fully in line with the stability considerations detailed in
the preceeding section. It remains to be seen whether the next order in the
expansion has a stabilising effect on the series
\cite{Bervillier:2005za}. \step

Within the ERG, parts of the cutoff space have been probed quantitatively to
order ${\cal O}(\partial^2)$
\cite{Tetradis:1993ts,VonGersdorff:2000kp,Canet:2002gs} and to order ${\cal
O}(\partial^4)$ \cite{Canet:2003qd}. Two observations have to be made in the
present context: first of all, the critical exponents $\nu$ and $\eta$ remain
bounded, similar to Fig.~1, in the parameter range considered. Furthermore,
they attain local extrema as functions of the cutoff indicating the existence
of a global boundary equivalent to those displayed in Fig.~1. The boundedness
of $\nu(R)$ within a given order of the derivative expansion is an important
ingredient in the convergence of the series. Secondly, the set of stable
flows, as identified through the optimisation, remains stable to higher
orders: typically, flows with regulators $R$ such that $\nu(R)$ to leading
order is in the vicinity of $\nu_{\rm opt}$ remain in the vicinity of the
local extrema of $\nu(R)$ even beyond leading order. This confirms the
validity of the underlying optimisation. For sufficiently stable flows, higher
order corrections remain quantitatively small, thus increasing the convergence
of the derivative expansion \cite{Litim:2001dt}.  \step

Finally, we point out that the qualitative differences beyond leading order
are also reflected in the explicit cutoff dependence of either flow. ERG flows
depend only on 'global' properties of the regulator $R$ through specific
momentum integrals $a_n(R)$ of the form
$$
a_n(R)\sim \int_0^\infty dy \frac{-y^{d/2+1-n}\, r'(y)}{[1+r(y)]^n}
$$
and similar \cite{Litim:2000ci}.  This follows from expanding the flow
\eq{ERG-Flow} in powers of $u'$ and $u'+2\rho u''$.  The coefficients $a_n(R)$
receive their main contributions for momenta $q^2\approx k^2$. Therefore,
small changes in the momentum behaviour of the cutoff $R\to R+\delta R$ induce
small changes in all coefficients $a_n(R)$. Furthermore, many different cutoff
functions $R$ can lead to equivalent sets of coefficients $a_n$. Hence, the
precise small- or large-momentum structure of $R$ is at best of subleading
relevance to the flow. In turn, Polchinski flows in a derivative expansion
depend both on 'global' and on 'local' characteristics of the regulator
$K(q^2/\Lambda^2)$, and in particular on its derivatives at vanishing
momenta. For example, the Polchinski flow for the wave function
renormalisation depends on the ratio $B(K)= K''(0)/K'(0)^2$, and the anomalous
dimension at criticality $\eta$ is even proportional to $B$ \cite{Ball:1995ji}
(see also \cite{Comellas:1998ep,Bervillier:2004mf,Bervillier:2005za}).  Small
modifications in the cutoff $K\to K+\delta K$ at small momenta can induce
large changes in $B$ including its sign, and, therefore, induce comparatively
large alterations in the flow and the physical observables.  These structural
differences can be seen as a further indication for the increased stability of
ERG flows as opposed to Polchinski flows.  \\[3ex]

\noindent
{\bf 7. Discussion and conclusions}\\[-1ex]

We compared several functional renormalisation group equations based on
Wilsonian cutoffs. The main structural differences between Polchinski flows
and ERG flows are due to the non-linearities of their right-hand sides. In the
literature, it has sometimes been argued that the simple non-linearities of
the Polchinski flow \eq{PRG} as opposed to those of ERG flows \eq{ERG} are a
benefit to the formalism. Here, we arrived at the opposite conclusion: the
non-linearities in ERG flows involve the inverse of $\Gamma^{(2)}_k$, and,
therefore, guarantee that the flow remains small in large regions of field and
momentum space. This structure implies that ERG flows are amiable to
systematic expansions ($e.g.$~in vertex functions) and allow for a
straightforward optimisation, since truncational variations in the flow are
suppressed in large parts of field space.  On the other side, the
non-linearities of the Polchinski flow appear to be algebraically simpler. The
price to pay is that the flow remains non-trivial in a larger domain of field
space, including the region of large fields. These differences in the locality
and stability behaviour favour the flows \eq{ERG} in particular for numerical
implementations.\step

The structural differences have been made explicit within a derivative
expansion. To leading order, critical exponents from the Polchinski flow are
scheme independent and, therefore, serve as a benchmark test for functional RG
flows in corresponding approximations. We have established the remarkable
result that the optimised ERG flow \eq{ERG-FlowOpt} and the Polchinski flow
\eq{PRG-Flow} have identical universal eigenvalues, for all $O(N)$ symmetric
scalar theories, for the leading and subleading critical exponents, and for
the asymmetric corrections-to-scaling exponent!  This equivalence is
non-trivial in that the corresponding flows \eq{ERG-FlowOpt} and \eq{PRG-Flow}
differ substantially, both in their structure and in their non-universal
scaling solutions.  We conjecture that this result extends to all universal
observables to leading order in the derivative expansion.\step

This equivalence, however, does not persist in an obvious manner beyond the
leading order, where universal observables from Polchinski flows depend
strongly on remaining unphysical parameters. This is in marked contrast to the
results from ERG flows which remain bounded, similar to the leading order. The
comparatively large cutoff sensitivity and the nontriviality of the Polchinski
flow for large fields -- a consequence of the non-linearities in \eq{PRG} --
require a better conceptual understanding before definite conclusions can be
drawn concerning its convergence properties. \step

For ERG flows, on the other hand, a coherent picture has emerged.  In given
truncations, an appropriate optimisation leads to an increased stability of
the flow.  The cutoff dependence of physical observables is, thereby, largely
reduced to a small range in the vicinity of the physical theory. This pattern
is established quantitatively within a derivative expansion, both to leading
order and beyond.  The comparatively weak cutoff sensitivity of optimised
flows and the triviality of the flow for large fields -- a consequence of the
non-linearities in \eq{ERG} -- are at the root for reliable applications of
the formalism to more complex theories including QCD and gravity.\\

\noindent
This work is supported by an EPSRC Advanced Fellowship.\\


\end{document}